\documentclass[10pt,twocolumn,twoside]{IEEEtran}
\usepackage{amsmath}
\usepackage{amssymb}
\DeclareMathOperator*{\argmax}{arg\,max}

\usepackage{graphicx}
\usepackage{epstopdf}
\usepackage{amsmath}
\usepackage{array}
\newcommand{\PreserveBackslash}[1]{\let\temp=\\#1\let\\=\temp}
\newcolumntype{C}[1]{>{\PreserveBackslash\centering}p{#1}}
\newcolumntype{R}[1]{>{\PreserveBackslash\raggedleft}p{#1}}
\newcolumntype{L}[1]{>{\PreserveBackslash\raggedright}p{#1}}
\usepackage[usenames]{color}
\usepackage{colortbl,booktabs}
\usepackage{tabularx}
\usepackage{multicol}
\usepackage{booktabs}
\usepackage[Symbol]{upgreek}
\usepackage{bm}
\usepackage{cite}
\usepackage{balance}
\usepackage{mathrsfs}
\usepackage{url}

\usepackage{threeparttable}
\usepackage{amsmath}
\usepackage{dcolumn}
\usepackage{multirow}
\usepackage{booktabs}
\usepackage{amsfonts}

\usepackage{subfigure}
\usepackage{algorithm}
\usepackage{algorithmic}

\begin{document}
	\title{Beam Selection for Wideband Millimeter Wave MIMO Relying on Lens Antenna Arrays}
	\author{\IEEEauthorblockN{Chenghao Feng,~\IEEEmembership{Student Member,~IEEE}, Wenqian Shen,~\IEEEmembership{Member,~IEEE}, and Jianping An,~\IEEEmembership{Member,~IEEE}\vspace{-10mm}}
        \thanks{
        This work was supported by the National Natural Science Foundation of China (NSFC) under Grant 61620106001.

		C. Feng, W. Shen, and J. An are with the School of Information and Electronics, Beijing Institute of Technology, Beijing 100081, China (e-mails: cfeng@bit.edu.cn, wshen@bit.edu.cn, and an@bit.edu.cn).
		}
	}
	\maketitle
	\begin{abstract}
Beamspace multi-input multi-output (MIMO) relying on lens antenna arrays can significantly reduce the number of radio-frequency chains in millimeter-wave (mmWave) communication systems through beam selection.
However, the beamforming gain is actually frequency-dependent in wideband mmWave MIMO systems.
This phenomenon is called beam squint, which will deteriorate the system's performance when traditional beam selection methods are used.
To solve this problem, we propose a wideband beam selection method for mmWave MIMO systems relying on lens antenna arrays.
Firstly, we select one beam with the maximal energy averaged over the whole band for each user and then we sequentially select the beams that contribute the most to the sum-rate.
Performance analysis of the proposed wideband beam selection method is also presented.
Numerical results show that the proposed method achieves higher sum-rate and energy efficiency compared with its traditional counterparts.
	\end{abstract}
	\IEEEpeerreviewmaketitle
	\section{Introduction}\label{S1}
	\IEEEPARstart{M}{illimeter}-wave (mmWave) communication has been identified as one of the key technology for 5G wireless communication systems due to its sufficient bandwidth \cite{JTSP_HRobert_OverviewMmwave,8198807}.
Massive multi-input multi-output (MIMO) technique is usually applied to compensate for the heavy path loss of mmWave signals through effective precoding.
 For the conventional fully digital precoding, each antenna is coupled to a dedicated radio-frequency (RF) chain.
 On account of the numerous antennas in mmWave MIMO systems, the fully digital precoding has extremely high hardware cost and power consumption. %
 To reduce the number of RF chains, a promising way is to use lens antenna arrays \cite{8291029,8141416}.
 Lens antenna arrays transform the spatial mmWave channel into sparse beamspace channel \cite{GLOBECOM_ASayeed_Beamspace}, meaning that only a part of beams carry much of the information.
 After beam selection, the number of RF chains is reduced at the cost of insignificant performance loss \cite{gao16bs}.

	To achieve this goal, \cite{GLOBECOM_ASayeed_Beamspace,gao16bs} exploited beam selection strategies that are suitable for narrow-band systems.
    The magnitude maximization (MM) method proposed in \cite{GLOBECOM_ASayeed_Beamspace} selects the focused-energy beams.
	Furthermore, the interference-aware beam selection (IA-BS) method proposed in \cite{gao16bs} reduces the interference among users to achieve high sum-rate.
    However, practical mmWave MIMO systems are wideband \cite{7402271}.
    When the angle-of-departure (AoD) is away from broadside of the lens antenna array, the required beams are frequency-dependent and act as a function of frequency.
    The phenomenon is called beam squint \cite{Brady2015Wideband,8422131}.
    Thus, the existing narrow-band beam selection methods are not directly applicable due to the effect of beam squint.
    The authors in \cite{8114345}  proposed a wideband beam selection method for the wideband mmWave systems using the spatial information in sub-6 GHz band.
    However, the effect of beam squint of wideband systems has not been considered in this paper.

	In this paper, we propose a wideband beam selection method to combat with the beam squint existing in wideband mmWave MIMO systems.
	Specifically, we select one beam with the largest energy averaged over the whole band for each user.
	Subsequently, we sequentially select the beams that contribute the most to the sum-rate until the target number of beams.
    Meanwhile, we derive the upper bound of the sum-rate gap between the fully digital structure and the proposed method.
    Numerical results show that the proposed wideband beam selection method outperforms the traditional methods in sum-rate and energy efficiency.
	
	\emph{Notation}:
	Lower-case and upper-case boldface letters denote vectors and matrices, respectively.
	$(\cdot)^{\rm{T}}$, $(\cdot)^{\text{H}}$, $(\cdot)^{-1}$ and $(\cdot)^{\dagger}$ denote the transpose, conjugate transpose, inverse and pseudo-inverse of a matrix, respectively.
    $tr(\cdot)$ represents the trace function.
    $\mathrm{Card} (\cdot)$ denotes the cardinality of a set.
    $|\cdot|$ and $\|\cdot \|$ denote the absolute value of a scalar and the norm of a vector.
    $\mathbf{A}_{\left(i,:\right)}$ and $\mathbf{A}_{\left(:,i\right)}$ are the $i$-th row and $i$-th column of the matrix $\mathbf{A}$, respectively.
	Finally, $\mathbf{I}_P$ denotes the identity matrix of size $P\times P$.
	\section{System Model And Channel Model}\label{S2}
In this section, we first introduce a wideband mmWave MIMO system relying on a lens antenna array.
Then we describe the wideband mmWave channel model, where the effect of beam squint is highlighted.
	\subsection{System model}\label{S2.1}
\begin{figure}[t]
		 \center{\includegraphics[width=0.45\textwidth]{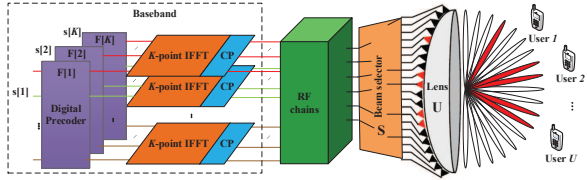}}
		\caption{Illustration of the wideband mmWave MIMO systems relying on a lens antenna array.}
		\label{narrowbeamspace}
	\end{figure}

We consider a multi-user wideband mmWave MIMO system with a base station (BS) serving $U$ single-antenna users, as shown in Fig. \ref{narrowbeamspace}.
The BS employs $N$ antennas and $N_{\mathrm{RF}}$ RF chains.
Different from the traditional narrow-band systems \cite{gao16bs}, the BS in the wideband system adopts the orthogonal frequency division multiplexing (OFDM) technology and the number of sub-carriers is $K$.
The BS transmits the signals $\mathbf{s}\left[ k \right] \in \mathbb{C}^{U \times 1}$ at each sub-carrier $k$ $\left(k=1,2,\cdots,K\right)$.
Note that the power of the transmit signal $\mathit{\mathbf{s}}\left[ k \right]$ is normalized as $\mathbb{E}\left( \mathbf{s}\left[ k \right] \mathbf{s}^\text{H}\left[ k \right] \right) = \mathbf{I}_U$.
Then the transmit signal is digitally precoded.
$\mathbf{F}\left[ k \right] \in \mathbb{C}^{N_{\mathrm{RF}} \times U}$ is the digital precoding matrix in the baseband.
It satisfies the power constraint $tr\left(\mathbf{F}\left[ k\right] \mathbf{F}^\text{H}\left[ k\right] \right)= \rho$, where $\rho$ is the transmit power.
Subsequently, $\mathit{K}$-point IFFTs are conducted to transform the precoded signals $\mathbf{F}\left[ k \right]\mathbf{s}\left[ k \right]$ into time domain. 
Afterwards, we add a cyclic prefix (CP) to the time-domain signal to deal with the inter symbol interference.
The beam selector $\mathbf{S} \in \mathbb{C}^{N \times N_{\mathrm{RF}}}$ selects $N_{\mathrm{RF}}$ antennas to be coupled with $N_{\mathrm{RF}}$ RF chains.
Each column of $\mathbf{S}$ has and only has one non-zero element `1' \cite{8322177}.
Thus, the received complex baseband signal $\mathbf{y}\left[ k \right] \in \mathbb{C}^{U \times 1}$ of $U$ users at the $k$-th sub-carrier is presented by
	\begin{align}\label{y_rec}
	\mathbf{y}\left[ k \right] = \mathbf{H}^{\text{H}}\left[ k \right]\mathbf{U}^{\text{H}}\mathbf{S}\mathbf{F}\left[ k \right]\mathbf{s}\left[ k \right] + \mathbf{n}\left[ k \right],
	\end{align}
where $\mathbf{H}\left[ k \right] = \left[ \mathbf{h}_1\left[ k \right],\mathbf{h}_2\left[ k \right], \cdots ,\mathbf{h}_U\left[ k \right] \right]\in \mathbb{C}^{N \times U}$ and $ \mathbf{h}_u\left[ k \right] $ is the spatial channel vector between the BS and the $u$-th user at the $k$-th sub-carrier.
$\mathbf{U}\in \mathbb{C}^{N \times N}$ is a DFT matrix because the lens performs as a spatial discrete fourier transform of the incident signals \cite{7582545}.
$\mathbf{n}\left[ k \right]\in \mathbb{C}^{U \times 1}$ is an additive Gaussian noise vector and $\mathbf{n}\left[ k \right] \sim \mathcal{CN}\left( \sigma^2\mathbf{I}_U \right)$ with $ \sigma^2 $ denoting the power of noise.

The beamspace channel $\tilde{\mathbf{H}}\left[ k \right] \in \mathbb{C}^{N \times U}$ is defined as $    \tilde{\mathbf{H}}\left[ k \right] = \mathbf{U}\mathbf{H}\left[ k \right] = \left[ \tilde{\mathbf{h}}_1\left[ k \right],\tilde{\mathbf{h}}_2\left[ k \right], \cdots ,\tilde{\mathbf{h}}_U\left[ k \right] \right]$.
Then we further define the reduced-dimension effective channel after beam selection as $\tilde{\mathbf{H}}_{\mathrm{r}}\left[ k \right] =\mathbf{S}^{\text{H}}\tilde{\mathbf{H}}\left[ k \right]= \tilde{\mathbf{H}}\left[ k \right]_{\left( n,: \right)\left( n \in \mathcal{B} \right)}$,
where $\mathcal{B}$ is the set of selected beams.
Thus, the digital precoder can be designed as $\mathbf{F}\left[ k \right] = {(\mathbf{\tilde{H}}^{\text{H}}_{\mathrm{r}}\left[ k \right])}^\dagger$ according to the widely used zero-forcing method.
It is worth noting that the signal processing in the baseband is frequency-dependent, while the beam selection is frequency-independent.
    \subsection{Channel model}\label{S2.2}
We will describe the wideband mmWave channel model in this sub-section.
The mmWave channel consists of a line-of-sight (LOS) propagation path and $L-1$ non-LOS (NLOS) propagation paths.
Therefore, the delay-$q$ channel tap for user $u$ can be expressed by \cite{8291029,Alkhateeb2016Frequency}
     \begin{align}\label{eqTHzhk1} 
      \mathbf{h}_{u,q} =  \sum_{\ell=0}^{ L - 1}&p_{\mathrm{rc}}\left( qT_s - \tau_{u,\ell} \right)\alpha_{u,\ell}\mathbf{a}\left( \phi_{u,\ell}^k\right),
     \end{align}
where $\ell=0$ denotes the LOS channel path, $p_{\mathrm{rc}}\left( \tau \right)$ denotes a raised-cosine filter performing as band-limited pulse-shaping function for $T_s$-spaced signaling evaluated at $\tau$ \cite{Alkhateeb2016Frequency}, $\tau_{u,\ell}$ is the delay, $\alpha_{u,\ell}$ is the complex gain.
$\mathbf{a}(\phi) = \frac{1}{\sqrt{N}}\left[ e^{1,e^{j2\pi\phi},\cdots, e^{j2\pi\phi\left(N-1\right)}}\right] ^{\rm{T}}$ \cite{GLOBECOM_ASayeed_Beamspace} is the array response vector when the typical uniform linear array is used at the BS.
$\phi_{u,\ell}^k = \frac{f_k}{c}d\mathrm{sin}\theta_{u,\ell}$ denotes the spatial AoD, where $f_k = f_c + \frac{B}{K}\left( k-1-\frac{K-1}{2} \right)$ is the frequency at sub-carrier $k$ with $f_c$ denoting the central frequency and $B$ denoting the bandwidth,
$c$ is the light speed, $d$ is the antenna spacing, and
$\theta_{u,\ell}$ is the physical AoD.
Note that the spatial AoD $\phi_{u,l}^k$ vary with frequencies, which is termed as beam squint \cite{8422131}.

Assuming that the CP is of length $N_Q$, the spatial channel vector of user $u$ at sub-carrier $\mathit{k}$ is
     \begin{align}\label{eqTHzhm}
     \mathbf{h}_{u}\left[ k \right] = \sum_{q=1}^{N_Q}\mathbf{h}_{u,q}e^{-j \frac{2\pi k}{K} q} = \sum_{\ell=0}^{L - 1} \beta_{u,\ell}\mathbf{a}\left( \phi_{u,\ell}^k \right),
     \end{align}
where $\beta_{u,\ell} = \alpha_{u,\ell}\sum_{q=1}^{N_Q}p_{\mathrm{rc}}\left( qT_s - \tau_{u,\ell} \right)e^{-j \frac{2\pi k}{K} q}$.
Due to the effect of beam squint, the beamspace channel $\tilde{\mathbf{H}}\left[ k \right]$ varies with the frequencies in different sub-carriers. However, the beam selection is frequency-independent.
Thus, the traditional narrow-band beam selection suffers from performance loss.
To solve this problem, we propose a wideband beam selection method in the following section.
\section{Wideband beam selection}\label{S3}
In this section, we propose a wideband beam selection method for mmWave MIMO systems.
We assume that the channel state information is known as prior information \cite{GLOBECOM_ASayeed_Beamspace,gao16bs}.
We also provide performance analysis of the proposed method.
\subsection{Proposed wideband beam selection method}\label{S3.1}
\begin{figure}[t]
    \centering
    \subfigure[ ]{
    \begin{minipage}[b]{0.35\textwidth}
    \includegraphics[width=1\textwidth]{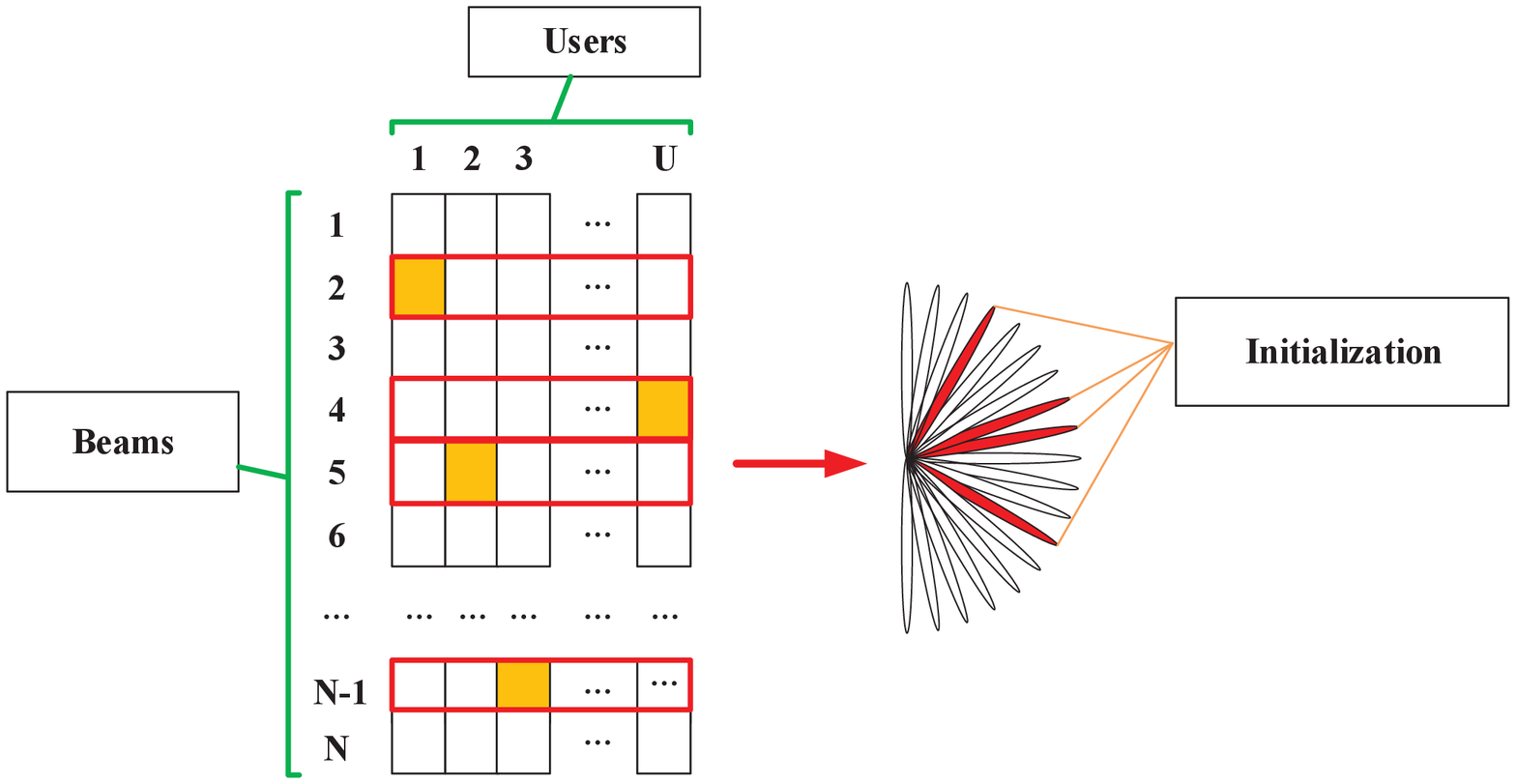}
    \end{minipage}
    }
    \subfigure[ ]{
    \begin{minipage}[b]{0.35\textwidth}
    \includegraphics[width=1\textwidth]{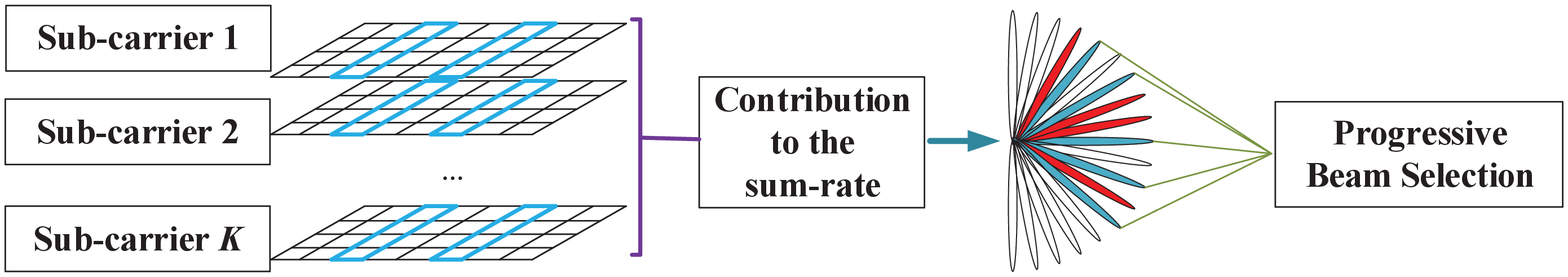}
    \end{minipage}
    }
    \caption{Illustration of the proposed wideband beam selection method: (a) initialization; (b) progressive beam selection}
    \label{algorithm}
    \end{figure}

Since that there exist $U$ users in the system, at least one beam should be allocated to each user \cite{GLOBECOM_ASayeed_Beamspace}.
To capture most of the channel energy over the whole band with limited number of RF chains through beam selection, we first propose to select one beam with the maximum energy averaged over the whole band for each user.
As shown in Fig. \ref{algorithm} (a), the 2D ``Beams v.s. Users" plane demonstrates the beam's energy averaged over $K$ sub-carriers.
For the remaining beams, we sequentially select the beams that contribute the most to the sum-rate as illustrated in Fig. \ref{algorithm} (b).
This step can be implemented in the following proposition 1.

\textit{\textbf{Proposition 1}}:
The problem that selecting the beam which contributes the most to the sum-rate is equal to finding the beam $b$ that satisfies
	\begin{align}\label{minbeam}
    \argmax_b\sum_{k=1}^{K}\mathrm{log}_2\left( 1 + \kappa_{b}\left[ k \right] \right),
	\end{align}
where $\kappa_{b}\left[ k \right] = \frac{tr\left( \mathbf{M}_b\left[ k \right] \right)  }{\left( tr\left(\mathbf{G}^{-1}\left[ k \right]\right) + 1 \right) \left( tr\left( \mathbf{G}^{-1}\left[ k \right] \right) - tr\left( \mathbf{M}_b\left[ k \right] \right) \right) }$,
$\mathbf{M}_b\left[ k \right] = \mathbf{g}_b\left[ k \right] \mathbf{G}^{-1}\left[ k \right] \mathbf{G}^{-1}\left[ k \right]\mathbf{g}_b^{\text{H}}\left[ k \right] \left( 1+ \mathbf{g}_b\left[ k \right]  \mathbf{G}^{-1}\left[ k \right] \mathbf{g}_b^{\text{H}}\left[ k \right]\right)^{-1}$,
$\mathbf{g}_b\left[ k \right] = \tilde{\mathbf{H}}\left[ k \right]_{\left( b,: \right)}$ and
$\mathbf{G}\left[ k \right] = \mathbf{\tilde{H}}_{\mathrm{r}}^{\text{H}}\left[ k \right] \mathbf{\tilde{H}}_{\mathrm{r}}\left[ k \right] + \delta\mathbf{I}$ with a small positive parameter $\delta$.
\begin{IEEEproof}
        Let $C_n$ be the sum-rate with the currently selected beam set $\mathcal{B}$ and $C_{n\prime}$ be the sum-rate with the updated beam set $\mathcal{B} \cup \{ b \}$.
        We can express the sum-rate of wideband mmWave MIMO systems as $C_n = U\sum_{k=1}^{K} \mathrm{log}_2\left(   1 + \frac{\xi}{tr\left( \mathbf{G}^{-1}\left[ k \right] \right)}  \right)$ and $C_{n\prime} = U\sum_{k=1}^{K} \mathrm{log}_2\left( 1 + \frac{\xi}{tr\left( \left(  \mathbf{G}\left[ k \right] + \mathbf{g}_b^{\text{H}}\left[ k \right] \mathbf{g}_b\left[ k \right] \right)^{-1} \right)} \right)$ \cite{gao16bs},
where $\xi = \rho / \sigma^2$ is the SNR.
Note that the inverse matrix of $\mathbf{G}\left[ k \right]$ is available for computation \cite{gao16bs}.
Thus, the sum-rate gain after selecting the beam $b$ can be expressed as
	\begin{align}\label{delta_C}
    \setlength{\abovedisplayskip}{1pt}
    \setlength{\belowdisplayskip}{1pt}
    C_{n\prime} - C_n  \overset{\left( a \right)}{=}  U\sum_{k=1}^{K} \mathrm{log}_2\left( 1 + \frac{tr\left( \mathbf{M}_b\left[ k \right] \right) \frac{ \xi } {tr\left( \mathbf{G}^{-1}\left[ k \right]\right) + \xi}  }{ tr\left( \mathbf{G}^{-1}\left[ k \right] \right) - tr\left( \mathbf{M}_b\left[ k \right] \right)}\right),
	\end{align}
where $(a)$ is obtained by the matrix inversion lemma, the property $tr\left( XY \right)=tr\left( YX \right)$ in matrix theory, and some mathematical operations.
The expression in the function $\mathrm{log}_2(\cdot)$ of (\ref{delta_C}) can be rewritten as $    1 + \frac{tr\left( \mathbf{M}_b\left[ k \right] \right)} {\left( \frac{1}{\xi} \cdot  tr\left( \mathbf{G}^{-1}\left[ k \right]\right) + 1 \right) \left( tr\left( \mathbf{G}^{-1}\left[ k \right] \right) - tr\left( \mathbf{M}_b\left[ k \right] \right) \right)}$.
Note that $\mathbf{G}\left[ k \right]$ is a diagonally dominant matrix because $\mathbf{h}_i^\text{H}\left[ k \right]\mathbf{h}_j\left[ k \right]\approx 0$ for $i\ne j$.
Then $tr\left( \mathbf{G}^{-1}\left[ k \right] \right) > 0$.
Besides, (\ref{delta_C}) is an increasing function for $tr\left( \mathbf{M}_b\left[ k \right] \right)$.
Thus, we set $\frac{1}{\xi}$ as $1$, which doesn't change the monotonicity of (\ref{delta_C}).
Then we select the beam $b$ to maximize (\ref{delta_C}), and (\ref{minbeam}) is acquired.
	\end{IEEEproof}
We summarize the proposed wideband beam selection method in \textbf{Algorithm 1}.
We first calculate the averaged energy in step 2.
Then we select one beam $b_u$ with maximal energy for each user in step 3 and 4.
We sequentially select the beams that contribute the most to the sum-rate until the target number of beams $N_{\mathrm{RF}}$ in step 8-10.
\begin{algorithm}[h]
\caption{Wideband beam selection method relying on a lens antenna array}
    \begin{algorithmic}[1]
        \REQUIRE ~~\\ 
        $\left\{ \tilde{\mathbf{H}}\left[ k \right] \right\}_{k=1}^K$ , $\mathcal{B}= \emptyset$, $N_{\mathrm{RF}}$ and $\mathcal{N}=\{1,2,\cdots,N\}$;
        \ENSURE
        \FOR{each user $u$}
        \STATE Calculate the averaged energy $\bar{\mathbf{h}} = \frac{1}{K}\sum_{k=1}^{K} \| \tilde{\mathbf{h}}_u\left[ k \right] \|^2$;
        \STATE Select the beam $b_{u} = \argmax_{b_{u}}\left| \bar{\mathbf{h}}\left(b_{u} \right) \right|$ , where $b_{u} \in \left\{1,2, \cdots, N \right\};$\
        \STATE Update $\mathcal{B}=\mathcal{B} \ \cup \ \left\{ b_{u}\right\} ;$\
        \ENDFOR
        \STATE $N_2=N_{\mathrm{RF}} - \mathrm{Card}\left( \mathcal{B} \right);$\
        \FOR {$i = 1 : N_2$ }
        \label{code:TrainBase:getc}
        \STATE Update $\mathbf{\tilde{H}}_r\left[ k \right] = \tilde{\mathbf{H}}\left[ k \right]_{\left( n,: \right)\left( n \in \mathcal{B} \right)}, \mathcal{N} = \mathcal{N} \backslash \mathcal{B};$
        \STATE Select the beam $b$ according to (\ref{minbeam});\
        \STATE Update $\mathcal{B}=\mathcal{B} \ \cup \ \left\{ b \right\};$\
        \label{code:TrainBase:pos}
        \ENDFOR
        \STATE Obtain the beamspace channel $\tilde{\mathbf{H}}_{\mathrm{r}}\left[ k \right] = \tilde{\mathbf{H}}\left[ k \right]_{\left( s,: \right)\left( s \in \mathcal{B} \right)}$
        \label{code:recentEnd}
    \end{algorithmic}
\end{algorithm}
\subsection{Performance analysis}\label{S3.2}
In this sub-section, we will analyze the sum-rate gap between the fully digital structure and the proposed wideband beam selection method.
We define the sum-rate of the fully digital structure and the proposed method as $C_D$ and $C_P$, respectively.

\textit{\textbf{Proposition 2}}:
When the SNR tends to be infinite, the sum-rate gap between the fully digital solution and the proposed method is upper bounded as
	\begin{align}\label{Prop 2}
    C_D \! - \! C_P \leq U\sum_{k=1}^{K} \mathrm{log}_2\left( 1 + \frac{ tr\left( \mathbf{M}\left[ k \right] \right) }{ tr\left( \mathbf{B}^{-1}\left[ k \right] \right) - tr\left( \mathbf{M}\left[ k \right] \right) }\right),
	\end{align}
    where $\mathbf{B}\left[ k \right] = \mathbf{\tilde{H}}_{\mathrm{r}}^{\text{H}}\left[ k \right] \mathbf{\tilde{H}}_{\mathrm{r}}\left[ k \right]$,
 $\mathbf{M}\left[ k \right] = \mathbf{P}\left[ k \right] \mathbf{B}^{-1} \left[ k \right] \mathbf{B}^{-1}\left[ k \right]$ $\mathbf{P}^{\text{H}}\left[ k \right]\left( \mathbf{I} + \mathbf{P}\left[ k \right]  \mathbf{B}^{-1}\left[ k \right] \mathbf{P}^{\text{H}}\left[ k \right]\right)^{-1}$, and
 $\mathbf{P}\left[ k \right]\! =\! \tilde{\mathbf{H}}\left[ k \right]_{\left( p,: \right)\left( p \in \mathcal{N} \backslash \mathcal{B} \right)}$.
\begin{IEEEproof}
The sum-rate of the fully digital structure is derived by selecting all remanent beams.
    Thus, the effective channel of fully digital structure is $\mathbf{\tilde{H}}_D\left[ k \right] = \begin{bmatrix} \mathbf{\tilde{H}}_{\mathrm{r}}\left[ k \right] \\ \mathbf{P}\left[ k \right] \end{bmatrix}$,
    and we have $\mathbf{\tilde{H}}_D^{\text{H}}\left[ k \right]\mathbf{\tilde{H}}_D\left[ k \right] = \mathbf{\tilde{H}}_{\mathrm{r}}^{\text{H}}\left[ k \right]\mathbf{\tilde{H}}_{\mathrm{r}}\left[ k \right] + \mathbf{P}^{\text{H}}\left[ k \right]\mathbf{P}\left[ k \right]$.
    Then the sum-rate of the fully digital solution is expressed by $C_D = U\sum_{k=1}^{K} \mathrm{log}_2\left(1 + \frac{\xi}{tr\left[ \left( \mathbf{\tilde{H}}_D^{\text{H}}\left[ k \right] \mathbf{\tilde{H}}_D\left[ k \right] \right)^{-1} \right]}\right)$.
      Subsequently, a proof similar to that of the Proposition 1 is adopted with the assumption that the inverse of $\mathbf{\tilde{H}}_{\mathrm{r}}^{\text{H}}\left[ k \right] \mathbf{\tilde{H}}_{\mathrm{r}}\left[ k \right]$ exists and the term $\mathbf{g}_b^{\text{H}}\left[ k \right] \mathbf{g}_b\left[ k \right]$ is replaced with $\mathbf{P}^{\text{H}}\left[ k \right] \mathbf{P}\left[ k \right]$ in $C_{n\prime}$.
      Then we derive the expression of the sum-rate gap as
      	\begin{align}\label{rate_gap}
     C_D - C_P = U\sum_{k=1}^{K} \mathrm{log}_2\left( 1 + \frac{tr\left( \mathbf{M}\left[ k \right] \right) \cdot \frac{ \xi } {tr\left( \mathbf{B}^{-1}\left[ k \right]\right) + \xi}  }{ tr\left( \mathbf{B}^{-1}\left[ k \right] \right) - tr\left( \mathbf{M}\left[ k \right] \right)}\right),
	\end{align}
 When $\xi$ tends to infinite, we have that $\lim_{\xi \to +\infty}\frac{\xi}{tr\left(\mathbf{B}^{-1}\left[ k \right]\right) + \xi} \!=\! 1$.
 Finally, (\ref{Prop 2}) is acquired.
  \end{IEEEproof}
   It is observed that the sum-rate gap increases as SNR becomes high.
 When the SNR tends to be infinite, the sum-rate gap asymptotically approaches the derived upper bound instead of increasing infinitely.
 \section{Numerical Results}\label{S4}
 In this section, we investigate the performance of the proposed wideband beam selection method in mmWave MIMO systems.
   Two aforementioned traditional methods, i.e., the MM method \cite{GLOBECOM_ASayeed_Beamspace} and the IA-BS method \cite{gao16bs}, are directly extended to wideband mmWave systems for comparison.
  	\begin{figure}[!htbp]
        \center{\includegraphics[width=0.4\textwidth]{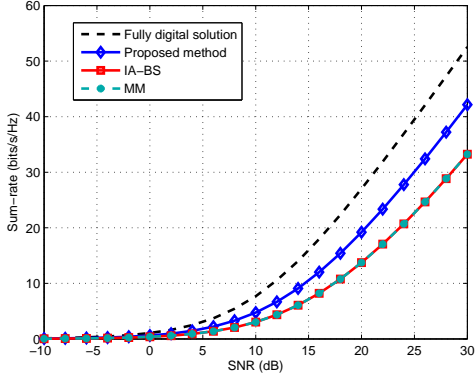}}
		\caption{Achievable sum-rate comparison. The proposed method outperforms two traditional methods.}
		\label{Beam_selection_vs_final}
	\end{figure}
 The system parameters for simulations are set as $N$=256, $K$=128, $L$=3, $U$=8, $N_{\mathrm{RF}}$=16 \cite{gao16bs}.
 $\mathrm{sin}\theta_{u,\ell}$ is i.i.d. and uniformly distributed over the interval $\left[-\frac{1}{2}, \frac{1}{2}\right]$ \cite{gao16bs}.
 We set $f_c$=28GHz and $B$=1.4GHz.

 We show the system sum-rate versus the SNR in Fig. \ref{Beam_selection_vs_final}.
 We observe that the proposed method outperforms traditional methods by about 5dB.
 Note that though IA-BS method eliminates the inter-user interference in each sub-carrier, the interference still exists over the wideband as a result of the effect of beam squint.
 Thus, the performance of MM method and IA-BS method is almost same.
	\begin{figure}[t]
		\center{\includegraphics[width=0.4\textwidth]{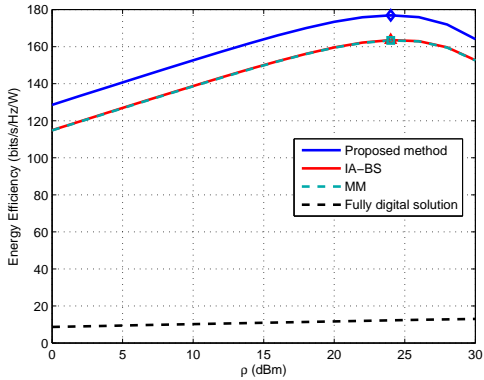}}
        \caption{Energy efficiency comparison. The proposed method outperforms two traditional methods.}
		\label{EE_vs_Ps}
	\end{figure}

The comparison about energy efficiency is shown in Fig. \ref{EE_vs_Ps}.
We calculate the energy efficiency as $ \eta = \frac{C}{\rho + N_{\mathrm{RF}}P_{\mathrm{RF}}}$ (bps/Hz/W) and the power of RF chains $P_{\mathrm{RF}}$ is typically set as 34.4mW \cite{GLOBECOM_ASayeed_Beamspace}.
The total transmit power $\rho$ ranges from 0dBm to 30dBm and the noise power $\sigma^2$ is -75 dBm \cite{Lin2016Energy} for all users.
We observe that the energy efficiency of the proposed method is much higher than that of traditional methods.

 Finally, we simulate three curves to verify the analytical results (\ref{Prop 2}), which include the sum-rate gap between the proposed method and the fully digital solution,
 the analytical expression of sum-rate gap (\ref{rate_gap}), and the derived upper bound (\ref{Prop 2}) when SNR goes infinite.
 We observe that the analytical results are consistent with the simulations.
 	\begin{figure}[t]
		\center{\includegraphics[width=0.4\textwidth]{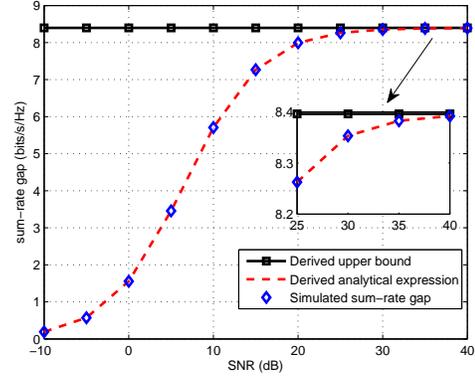}}
        \caption{Sum-rate gap versus SNR. The sum-rate gap improves as SNR increases and it infinitely approaches the upper bound.}
		\label{analytical}
	\end{figure}
\section{Conclusion}\label{S5}
    The beam squint existing in wideband mmWave MIMO systems deteriorates the system performance when traditional narrow-band beam selection methods are used.
    In this paper, we proposed a wideband beam selection method to solve this problem.
    Specifically, we selected one beam with the maximum energy over the wideband for each user and then we sequentially selected the beams that contribute the most to the sum-rate.
    We also derived the upper bound of the the sum-rate gap between the fully digital solution and the proposed method when the SNR goes infinite.
    Numerical results verified the superior performance of the proposed method.
	\bibliographystyle{IEEEtran}
	\bibliography{IEEEabrv,Refference}
\end{document}